\documentclass[11pt,%twocolumn,dvips
]{article}
\usepackage[]{graphicx}
\usepackage{makeidx}
\newcommand{\mathsym}[1]{{}}

\newcommand{\bi}{\bibitem}
\newcommand{\be}{\begin{eqnarray}}
\newcommand{\ee}{\end{eqnarray}}
\newcommand{\rar}{\rightarrow}

\usepackage[]{caption}
\def\gsim{\mathrel{\raise.3ex\hbox{$>$\kern-.75em\lower1ex\hbox{$\sim$}}}}
\def\lsim{\mathrel{\raise.3ex\hbox{$<$\kern-.75em\lower1ex\hbox{$\sim$}}}}

\begin{document}

\begin{center}
{\bf \large{Screening effects in plasma with charged Bose condensate
}} \\ \vspace{0.5cm}
{\it Alexander D. Dolgov \footnote{dolgov@fe.infn.it}$^{a,b,c}$, 
Angela Lepidi \footnote{lepidi@fe.infn.it}$^{a,b}$,
 and Gabriella Piccinelli \footnote{gabriela@astroscu.unam.mx}$^{d,a}$}
\\\vspace{0.3cm}
$^a$ Istituto Nazionale di Fisica Nucleare, Sezione di Ferrara, 
I-44100 Ferrara, Italy \\
$^b$ Dipartimento di Fisica, Universit\`a degli Studi di Ferrara, 
I-44100 Ferrara, Italy \\
$^c$ Institute of Theoretical and Experimental Physics, 113259 Moscow, Russia \\
$^d$ Centro Tecnol\'ogico, FES Arag\'on, Universidad Nacional Aut\'onoma de  M\'exico, 
Avenida Rancho Seco S/N, Bosques de Arag\'on, Nezahualc\'oyotl, 
Estado de M\'exico 57130,  M\'exico

\begin{abstract}
Screening of Coulomb field of test charge in plasma with Bose condensate of
electrically charged scalar field is considered. It is found that the screened 
potential contains several different terms: one decreases as a power of distance
(in contrast to the usual exponential Debye screening), some other oscillate with
an exponentially decreasing envelope. Similar phenomenon exists for fermions (Friedel 
oscillations), but fermionic and bosonic systems have quite different features.
Several limiting cases and values of the parameters are considered and the 
resulting potentials are presented.
\end{abstract}
\end{center}

%----------------------------------------------------------------------------------------------------------------------------------------
\section{Introduction}
%----------------------------------------------------------------------------------------------------------------------------------------

It is well known that an electric charge, $Q$, in plasma is screened according 
to the Debye law, so the long-ranged Coulomb field is transformed into the
Yukawa type potential (see e.g.~\cite{landau-9,Kapusta:1989tk}):   
\be
U(r) = \frac{Q}{4 \pi r}\,\rar \frac{Q\,\exp (-m_D r)}{4\pi r}\,,
\label{debye}
\ee
where the Debye screening mass, $m_D$, is expressed through the plasma temperature and
chemical potentials of the charged particles, see below 
eqs. (\ref{Pi-0-rel},\ref{Pi-0-nonrel}). Physical 
interpretation of this result is evident: test charge polarizes plasma around, 
attracting opposite charge particles and thus the electrostatic field drops down
exponentially faster than in vacuum. Formally the Debye screening appears from a pole
at purely imaginary $k$ in the photon propagator in plasma, $(k^2 + \Pi_{00})^{-1}$,
where $\Pi_{00}$ is the time-time component of the photon polarization operator.

By an evident reason the screening effects were studied historically first in 
fermionic i.e. in electron-proton and in electron-positron plasma. For degenerate
fermionic plasma another and quite striking screening behavior was found. Namely
the screened potential drops down as a power of distance, $1/r^3$ in 
non-relativistic case and $1/r^4$ in relativistic case multiplied by an 
oscillating function, $\cos (k_Fr)$ or $ \sin (k_F r)$,
where $k_F$ is the Fermi momentum. This phenomenon is called
Friedel oscillations~\cite{friedel,fetter}. Usually it is prescribed to a sharp 
(non-analytic) cut-off of the
Fermi distribution of degenerate electronic plasma at $T=0$
but maybe it is better to say
that the effect is related to the logarithmic singularity of the photon polarization
operator $\Pi_{00} (\omega =0, k)$. This type of screening is discussed in 
sec.~\ref{s-friedel} both in non-relativistic and relativistic cases for arbitrary,
not necessarily zero, temperature. 
 
Plasma with charged bosons attracted attention much later, both for pure scalar 
electrodynamics, for a review see e.g. ref.~\cite{Kraemmer:1994az}, 
or for quark-gluon plasma~\cite{Kapusta:1989tk,bellac,yagi}. 
Surprisingly until last year
the impact of possible Bose condensate of charged fields on the photon
polarization operator was not considered. Only recently an investigation of 
plasma with Bose condensate of charged scalars was 
initiated~\cite{gr-1}-\cite{gaba-ros2}. 
It was found that in presence of Bose condensate the screened potential behaves
similarly to that in fermionic case, i.e. the potential oscillates, exponentially
decreasing with distance~\cite{gr-2,dlp}. This effect, however, 
in contrast to Friedel oscillations, does not come from the logarithmic 
branch point singularity in $\Pi_{00}$ 
but from the pole in the photon propagator at complex (not purely imaginary) value
of $k$. It was shown that the polarization operator contains infrared singular term
$\Pi_{00} \sim 1/k^2$~\cite{dlp,gr-3} which shifts the pole position
from imaginary axis (as in
Debye case) to a point with non-zero real and imaginary parts.

At non-zero temperature the polarization operator has another infrared singular
term $\sim 1/k$. This term is odd with respect to the parity transformation, 
$k \rar -k$ and, as a result, the potential acquires the term which decreases
as a power of distance but does not oscillate.
Moreover, the polarization operator has logarithmic singularity as
in the fermionic case and this singularity also generates an oscillating potential
similar to the Friedel one. It is interesting that the screened potential is a
non-analytic function of the electric charge $e$. In particular in certain limit it may 
be inversely proportional to $e$, despite being calculated in the lowest order in $e^2$.

Another oscillatory, exponentially damped, behavior of the potential
between static charges have been reported in the literature: 
it was  argued~\cite{Diaz} that in
nuclear matter at high densities and low temperatures, the Debye pole 
acquires a
non-zero real part and so the screened potential oscillates 
(see also ref.~\cite{Mu}). 
These {\it Yukawa oscillations} are short-ranged oscillations and 
fade away with distance faster, as compared to the Friedel oscillations.

In this paper we further analyze the asymptotics of the screening 
effects arising in bosonic and fermionic plasma.
In particular we have taken into account all the contributions, including the ones 
from the logarithmic singularities in the photon polarization tensor, and 
considered different limiting cases.

The content of the paper is the following.
In sec. \ref{pol-op-sec} we reproduce our results for the photon polarization 
operator in plasma with charged Bose condensate. 
Fermionic Friedel oscillations in non-relativistic and relativistic cases 
both for T=0 and $T\neq 0$ are considered in sec. \ref{friedel-sec}.
There we reproduce already established results but using different techniques.
In sec. \ref{bos-sec} we calculate screening in bosonic plasma, taking into account 
the contributions from the 
poles in the complex $k$-plane, from the integral along the imaginary axis, and 
from the logarithmic branch cuts. 
The last part has never been done before. We consider several conditions, 
in particular bosons with or without  condensate and eventually even in absence 
of fermions.
Finally, in sec. \ref{conclusion} our conclusions are presented.

%----------------------------------------------------------------------------------------------------------------------------------------
\section{Polarization operator of photon in medium \label{s-polarization}}
\label{pol-op-sec}
%----------------------------------------------------------------------------------------------------------------------------------------

We confine ourselves to the lowest order in the electromagnetic coupling, $e^2$. The 
photon polarization operator, $\Pi_{\mu\nu} (\omega, k)$, in this approximation is
well known, see e.g. books~\cite{Kapusta:1989tk,bellac}. For the calculation of the
latter either imaginary or real time methods are used. However, the result
can be obtained in a simpler way~\cite{dlp} 
just including into the photon Green's function the
effects of medium, namely, taking not only expectation value of the time ordered
product of $\langle A_\mu (x) A_\nu(y) \rangle$ over vacuum but add also the matter
states with the weight equal to the particle distribution, $f_j(q)$, where $j$
denotes the particle type and $q$ is the particle momentum. 
The resulting expressions, found in many works - see e.g. \cite{dlp} and 
reference therein - are the following:
	\begin{eqnarray}
	\label{Pi-B}
	\Pi_{\mu\nu}^{B} (k) 
	&=& e^2 \int \frac{d^3q}{(2 \pi)^3 E} 
	\left[ f_B (E,\mu) + \bar f_B (E,\bar\mu) \right] \times
\cr\cr
	&&\left[  \frac{1}{2} \, \frac{(2q - k)_\mu (2q - k)_\nu}{(q - k)^2 - m_B^2}
	+\frac{1}{2} \, \frac{(2q + k)_\mu (2q + k)_\nu}{(q + k)^2 - m_B^2} 
 -   g_{\mu\nu}\right], \\
%	\cr\cr
	\Pi_{\mu\nu}^{F} (k)
	&=& 2 e^2 \int \frac{d^3q}{(2 \pi)^3 E} 
	\left[ f_F (E,\mu) + \bar f_F (E,\bar\mu) \right] \times
	\cr\cr 
	&& \left[
	\frac{q_\nu (k+q)_\mu - q^\rho k_\rho g_{\mu\nu} + q_\mu (k+q)_\nu}{(k+q)^2-m_F^2} 
	+ \frac{q_\nu (q-k)_\mu + q^\rho k_\rho g_{\mu\nu} + q_\mu (q-k)_\nu}{(k-q)^2-m_F^2}
	\right],
\label{Pi-F}
	\end{eqnarray}
where $k=[\omega, {\bf k}]$ and $q=[E, {\bf q}]$ are four momenta of photon and charged 
particles living in plasma,
$E=(q^2 + m_{B,F}^2)^{1/2}$, with
$m_{B,F}$ being either mass of charged bosons or fermions, 
$\Pi_{\mu\nu}^B$ and $\Pi_{\mu\nu}^F$ are respectively the contribution to  the
polarization tensor from bosons and fermions,
$\mu$ and $\bar\mu$ are chemical potentials for particles and antiparticles.
Chemical potentials for bosons and fermions are generally unequal, moreover, the
chemical equilibrium is not necessarily maintained and $\mu + \bar\mu \neq 0$. Though
in what follows we present all the results for $\mu + \bar\mu =0$, it is straightforward
to generalize them to arbitrary $\mu$ and $\bar\mu$.

In kinetic equilibrium the distribution functions take the form:
\be
f_{B,F} = \frac{1}{\exp\left[(E-\mu)/T\right] \pm 1},
\label{f-equil}
\ee
where the signs $"+"$ and $"-"$ stay for fermions and bosons respectively.
In the case that the boson chemical potential is equal to its maximum allowed value
$\mu = m_B$ (or $\bar\mu = m_B$) the formation of Bose condensate is possible and in 
equilibrium the Bose distribution function  acquires an additional
term describing accumulation of  bosons in zero momentum mode:
\be
f_B  = C \delta^{(3)} ({\bf q}) +
\frac{1}{\exp\left[(E-m_B)/T\right] \pm 1}\,,
\label{cond}
\ee
where $C$ is a constant parameter describing the amplitude of the condensate.

The screening of test charge in the static case is determined by the zero frequency value
of $\Pi_{00} (0,k)$. We assume that plasma is homogeneous and isotropic, 
so the polarization
tensor depends only upon the magnitude of vector ${\bf k}$ but not on its direction. The 
corresponding expressions can be easily read from eqs.~(\ref{Pi-B},\ref{Pi-F}):
	\begin{eqnarray}
	\label{Pi-00-B}
	\Pi_{00}^B  (0,  {k}  ) &=&
	\frac{e^2}{2\pi^2} \int_0^\infty  \frac{dq \,q^2}{E_B} 
\left[f_B(E_B,\mu_B)+ \bar f_B(E_B,\bar\mu_B)\right] 	
	\left[1+\frac{ E^2_B}{ k q} \ln \bigg| \frac{2q +k}{2q - k} \bigg| \right], 
%\label{Pi-00-B}
\\
\Pi_{00}^F  (0,  {k}  ) &=&	\frac{e^2}{2\pi^2} \int_0^\infty \frac{dq\,q^2}{E_F} 
\left[f_F(E_F,\mu_F)+ \bar f_F(E_F,\bar\mu_F)\right] 	
	\left[2+\frac{ (4E_F^2 - k^2)}{2 k q} \ln \bigg| 
\frac{2q +k}{2q - k} \bigg| \right].
\label{Pi-00-F}	
\end{eqnarray}
In what follows we will omit the first argument in the polarization tensor, i.e.
write $\Pi_{00} (0,k) \equiv \Pi_{00} (k)$.

Evidently the first (condensate) term in $f_B$ gives rise to the quadratic infrared 
singularity $\Pi_{00}\sim 1/k^2$, as found in refs.~\cite{dlp,gr-3}. 
At non-zero temperature the pole singularity of the Bose distribution at $q=0$ 
leads to an additional infrared pole $\sim 1/k$ in the 
polarization tensor of photons~\cite{dlp}.
Thus at low values of the photon momentum $\Pi_{00}$ can be expanded as~\cite{dlp}:
\begin{eqnarray}
\Pi^B_{00}(0,k) &=&  
e^2\left[ h(T) +
+\frac{m_B^2 T}{2k} +
\frac{1}{(2\pi)^3} \, \frac{C}{m_B}
	\left( 1 +  \frac{4m_B^2}{ k^2} \right)  \right],
\label{low-k-Pi-00}
\end{eqnarray}
where the function $h(T)$ is independent of $k$ and has the limiting values:
	\begin{eqnarray}
	h(T) &=& \Biggl\{ 
	\begin{array}{c} T^2/3  \hspace{4.5cm} \rm{(high \, T)}
	\\ \zeta(3/2) (m_B T^3)^{1/2} / (2\pi)^{3/2} \hspace{1.3cm} \rm{(low \, T)}
\label{h-of-T}	
\end{array}.
	\end{eqnarray}
The low $T$ limit of the function $h(T)$ is however always sub-dominant with 
respect to the second term in eq. (\ref{low-k-Pi-00}) which comes
from the logarithmic term in eq. (\ref{Pi-00-B}). 

In the expression of the photon polarization tensor written above the 
singularities of $\Pi_{00}$
due to pinching of the integration contour by the poles of $f_B(E_B, m_B)$
and the logarithmic branch point in the integrand of 
eq.~(\ref{Pi-00-B}) are not taken into account.
It will be done below in sec. \ref{bos-sec}

The contribution of fermions into the polarization tensor is not infrared singular,
so it is convenient to present the latter as 
\be
\Pi_{00}^F (k) = \Pi_{00}^F (0) + \left[\Pi_{00}^F (k) - \Pi_{00}^F (0)\right]\,,
\label{Pi-of-k-0}
\ee
where 
\be
\Pi_{00}^F (0) = \frac{e^2}{\pi^2}\,\int\frac{dq}{E} (f+\bar f) (q^2 + E^2) \,.
\label{Pi-of-0}
\ee

In the case of relativistic fermions with non-zero chemical potential, $\mu$, 
the zero momentum limit  of $\Pi_{00}^F$ is~\cite{fradkin}:
\begin{eqnarray}
\Pi_{00}^F( 0) = e^2\left( \frac{T^2}{3} + \frac{\mu^2}{\pi^2} \right).
\label{Pi-0-rel}
\end{eqnarray}
This expression is valid in the limit $m_F\ll \mu$ and $m_F \ll T$,
while in non-relativistic case for positive $(\mu-m)$ and small $T$ 
we find:
\be
\Pi_{00}^F( 0) = \frac{\sqrt{2} e^2 m_F^{3/2} (\mu-m_F)^{1/2}}{\pi^2}
- \frac{e^2 T^2}{12\sqrt{2}}\,\left(\frac{m_F}{\mu-m_F}\right)^{3/2}+ ... \, .
\label{Pi-0-nonrel}
\ee
If $\mu<m$, the polarization tensor is exponentially suppressed, 
$\Pi_{00} \sim \exp[-(m-\mu)/T]$. For the Debye mass we find the well known 
non-relativistic result:
\begin{eqnarray}
m_D^2 = \frac{e^2 n_F}{T}.
\label{m-D-nonrel}
\end{eqnarray}
Here, as above in the bosonic case, the singularities of $\Pi^F_{00}$
due to logarithmic branch point in integral (\ref{Pi-00-F})
are not included. For that  see the next section.

The potential of a test charge, $Q$, modified by the plasma screening effects
is given by the Fourier transform of the photon propagator in plasma:
\begin{eqnarray}
U(r) = \frac{Q}{(2\pi)^3} \int \frac{d^3 k \exp (i {\bf k r})}{k^2 + \Pi_{00} (k)} =
\frac{Q}{2\pi^2}\int_0^\infty \frac{dk k^2}{k^2 + \Pi_{00} (k)}\,
\frac{ \sin (kr)} {kr} =
\frac{Q}{2\pi^2 r} {\cal I}m \int_0^\infty \frac{dk k\,e^{ikr}}
{k^2 + \Pi_{00} (k)}\,.
\label{U-of-r}
\end{eqnarray}
Usually the integrand in eq. (\ref{U-of-r}) is an even function on $k$ and
the integration along the line of positive real $k$ can be transformed into
the contour integral in the upper complex k-plane. 
%for $\exp(ikr)$ and in the lower plane for $\exp (-ikr)$. 
However, in the case of bosons with $\mu_B = m_B$
the polarization operator contains and odd term $m_B^2 T/2k$, eq. (\ref{low-k-Pi-00}), 
and the usual contour transformation is not applicable. So we express integral
(\ref{U-of-r}) through the integral along imaginary upper $k$-axis plus 
contribution of singularities in the upper $k$-plane. If $\Pi_{00}$ is an even 
function of $k$ and  $(k^2+\Pi_{00})^{-1}$ is regular on the imaginary $k$-axis,
the imaginary part of the integral along the imaginary axis vanishes.
If the integrand has a pole at positive imaginary $k=ik_D$, i.e.
\be
-k_D^2 + \Pi_{00} (ik_D) = 0,
\label{k-D}
\ee
this poles contributes into the integral as $i\pi \delta (k-ik_D)$ and gives
rise to the usual exponential Debye screening.
If $\Pi_{00}$ contains an odd part, the integral along imaginary $k$ axis
gives a contribution to the potential which decreases only as power of
distance~\cite{dlp}. 

There may also be poles at complex $k=k_p$, when both real and imaginary parts of
$k_p$ are non-zero. Such poles have been found for plasma with charged Bose 
condensate~\cite{gr-2,dlp}. They produce oscillating behavior superimposed
on the exponential decrease of the potential. It was argued~\cite{Diaz} 
that complex poles also exist in plasma of strongly interacting particles 
(pions and nucleons) and in QCD plasma.

There are also logarithmic singularities of $\Pi_{00}(k)$ at some non-zero 
${\cal I}m\,k$ and the integrals along the corresponding cuts also produce 
oscillations in the screened potential but the exponential cut-off is much weaker,
it is proportional to temperature and for zero $T$ it becomes a power law one.
For fermions this effect, called Friedel oscillations,
is known for a long time~\cite{friedel,fetter}, while for bosons 
a similar phenomenon has not been studied before.

%----------------------------------------------------------------------------------------------------------------------------------------
\section{Friedel oscillations in fermionic plasma \label{s-friedel}}
\label{friedel-sec}
%----------------------------------------------------------------------------------------------------------------------------------------

We consider here the Friedel oscillations in fermionic plasma. The non-relativistic
case is discussed in ref.~\cite{friedel,fetter,kap-toi}, 
both at zero and non-zero temperatures. 
The relativistic case was studied in~\cite{kap-toi}.
In what follows all four cases are presented, considered in somewhat different way.

Singularities of $\Pi_{00} (k)$ in the complex $k$-plane appear when the singular points
of the integrand in eq.~(\ref{Pi-00-F}) in the complex $q$-plane pinch the contour of 
integration or coincide with the integration limit at $q=0$. The usual calculation
is done at zero temperature when the fermion distribution tends to theta-function
and hence the integral over $dq$ goes from zero to the Fermi momentum, $q_f$. The 
singularity in $\Pi_{00}^F$ appears when the branch points of the logarithm at 
$k=\pm 2q$ move to the integration limit at $q=q_F$. In more general case of arbitrary
temperature the integrand is a smooth function of $q$ and integration goes up to
infinity. The integrand has two kinds 
of singularities. First, there are poles in the distribution function $f_F$ which are 
situated at
\be
q_n^2 = \left[ \mu \pm i\pi T(2n+1)\right]^2 -m_F^2 ,
\label{q-n}
\ee
where $n$ runs from 0 to infinity.

\begin{figure}[htbp]
	\begin{center}
	 \includegraphics[scale=0.50]{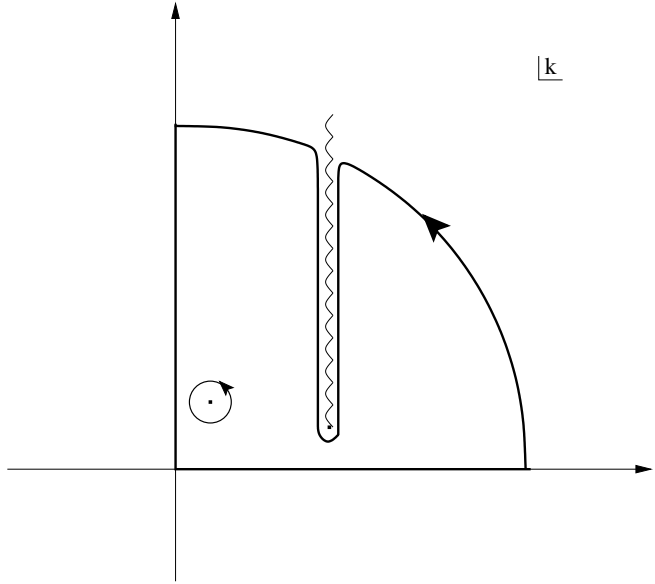}	
	\caption{{\it Contour of integration in complex k-plane.}}
	\label{fig_contour}
	\end{center}
	\end{figure}

The second type of singularities are branch points of the logarithm at
\be
q_b = \pm k/2\,.
\label{q-b}
\ee
The singularities of $\Pi_{00} (k)$ are situated at such $k_n$ for which $q_n$
and $q_b$ coincide, $q_n=q_b$ and
the poles, $q_n$, and branch points, $q_b$ approach the integration contour in $q$-plane
from the opposite sides. Since, according to the discussion in the previous section and
eq. (\ref{U-of-r}), we consider $k$ in the first quadrant of the complex $k$-plane,
only the singularities with ${\cal R}e k \geq 0$ and ${\cal I}m k \geq 0$ contribute to
the asymptotics of the potential, i.e.
\be
k_n = 2q_n = \left[ \left( \mu + i\pi T (2n+1)\right)^2 - m_F^2\right]^{1/2}.
\label{k-2qn} 
\ee
Symbolically the integral in the r.h.s. of eq.~(\ref{U-of-r}) can be written as a 
sum of three contributions:
\be 
I_0 = \int_0^{\infty} [i dk ] + 2\pi i \sum [ Res] + 
\sum_n \int_{k_n}^{k_n +i\infty} \Delta \,,
\label{I-0}
\ee
where the first integral goes along the positive imaginary axis in $k$-plane,
the second one is the sum of the residues of the poles on the integrand (if the poles
are on the imaginary axis, only a half of the residue is to be taken), and
the third term is the integral of the discontinuity over the branch line of the 
logarithmic singularity of $\Pi_{00}(k)$. The integration contour in complex $k$-plane is schematically
depicted in Fig. \ref{fig_contour}, where only one pole and one branch-cut are included. 

Before calculating the singular part of $\Pi_{00}$ let us first note that we are 
interested only in singularities in the first quadrant in $k$-plane and thus only
contribution from $-\ln |2q - k|$ should be taken. Since the absolute value of the
argument can be written as the limit of $\epsilon \rar 0$ of
$|2q-k| = [(2q -k)^2 + \epsilon^2 ]^{1/2} $, the logarithmic contribution into
$U(r)$ is given by
\be
\ln \bigg| \frac{k+2q}{k-2q }\bigg| \rar - \ln \bigg|k- 2q \bigg|  =
-\left[ \ln (k-2q + i\epsilon) + \ln (k-2q - i\epsilon)
\right]/2 \rar -\ln (k-2q - i\epsilon) /2.
\label{logs}
\ee
 
The singular part $\Pi_{00}^{(n)}$ near $k_n$ can be determined as follows. The integral
along the contour squeezed 
between $q_n$ and $q_b$ is equal to the residue of the integrand at the pole multiplied
by $2\pi i$ plus a regular part at $k=k_n$. The pole term near $q = q_n + z$
is equal to
\be
\frac{1}{\exp \left[ ({E_n -\mu})/T\right] + 1} = -\frac{E_n T}{z q_n}.
\label{pole}
\ee
The residue in the pole gives the singular term in
$\Pi_{00}$ equal to:
\be
\Pi_{00}^{(n)} (k) = \frac{ie^2 T}{4\pi k}\,(4E_n^2 - k^2)\,
\ln (k-2q_n - i\epsilon)\,, 
\label{Pi-n}
\ee 
where $q_n$ is the pole position given by eq. (\ref{k-2qn}) and 
$E_n = \sqrt{q_n^2+ m^2} $.
We have neglected here the contributions of
antiparticles assuming that the chemical potential is sufficiently large.
The discontinuity of $\Pi_{00} $ at the branch line 
$k = 2q_n + i y$, where $y$ runs from zero to infinity, is equal to
\be
\Delta \Pi_{00}^{(n)} =  \Pi_{00}^{(n)+} -  \Pi_{00}^{(n)-} =
- \frac{e^2 T (4 E_n^2 - k^2)}{ 2k}\,,
\label{Delta-Pi}
\ee
where upper index $"+"$ or $"-"$ indicate that the value of $\Pi_{00}$ is
taken on the right or the left hand side of the cut. 

The contribution of this discontinuity into the asymptotic 
behavior of $U(r)$, eq.~(\ref{U-of-r}), is equal to:
\be
U_{n} (r) = \frac{Q}{2\pi^2 r }\, {\cal I}m\,\int_0^{\infty} 
\frac{i dy\, k \exp (-y r + 2i q_n r) \left( - \Delta \Pi_{00}\right) }
{\left[ k^2 + \Pi_{00}^{(n)+}(k)\right]
\left[ k^2 + \Pi_{00}^{(n)-}(k)\right]}\,.
\label{U-cut-n}
\ee
Here $k = 2q_n + iy$. For fermionic plasma
we can neglect $y$ in comparison with $q_n$ because in the limit of large 
distances $y\sim 1/r$. However in the bosonic case a non-vanishing contribution
comes from sub-dominant in $y$ terms, see below.

Below we consider separately, firstly, relativistic and, secondly, 
non-relativistic cases. In 
relativistic limit $E_n = q_n$ and the factor in front of logarithm, eq. (\ref{Pi-n}), 
and discontinuity (\ref{Delta-Pi}) vanish at the branch point and 
the discontinuity becomes purely
imaginary in the leading order,
$\Delta \Pi_{00}^{(n)} = i e^2 T y$.
This leads to a faster decrease of the screened potential in 
comparison with non-relativistic case, $1/r^4$ instead of $1/r^3$, and to the 
change of phase, $\sin (2\mu r)$ instead of $\cos (2\mu r)$.

In relativistic case, when $m\ll T$ but $\mu$ may be large, the poles are 
situated at:
\be
E_n = q_n = \mu \pm i\pi T (2n+1)\,.
\label{q-n-rel}
\ee
Since $|k|^2 > 4 |q_n|^2 > 4(\mu^2- m_F^2)$, then for sufficiently large $\mu$,
$\mu>m_F$, and low $T$ we can neglect 
$\Pi_{00}\sim e^2 \mu^2$ in the denominator in comparison with $4q_n^2$  
and obtain:
\be
U_{n} (r) = \frac{Q e^2 T}{ 16 \pi^2 q_n^3 r^3}\, {\cal I}m\, e^{2iq_n r}=
\frac{Q e^2 T}{ 16 \pi^2 q_n^3 r^3}\,\sin(2\mu r) e^{-2\pi (2n+1) T r}.
\label{U-n-rel}
\ee
For non negligible $T$ the dominant term is that with $n = 0$ and though it decreases
exponentially, the power of the exponent may be much smaller than the standard
one, eq. (\ref{debye}) with $m_D= e\mu/\pi$, as follows from eq. (\ref{Pi-0-rel}).

At small $T$ the result is proportional to the temperature and thus 
formally vanishes at $T=0$.
However, at small $T$ the total contributions of the branch points diverges as 
$1/T$, so summing up all $U_n$ we find
\be
U_{cut} = \sum_{n=0}^\infty  U_n = 
\frac{e^2 Q T}{16\pi^2 r^3 \mu^3 }\,\frac{\sin (2\mu r)\,
\exp(-2\pi rT)}{1 - \exp (-4\pi r T)}\,.
\label{U-cut}
\ee
For $T\rar 0$ and large $r$ we can take $q_n =\mu$ because the effective $n$'s are of the
order of $n_{eff} \sim 1/(4\pi r T )$ and $nT \sim 1/r \ll \mu$.

For very small $T$ such that $rT \ll 1$ we obtain: 
\be
U_{cut} = \frac{e^2 Q }{64\pi^3} \frac{\sin (2\mu r)}{r^4 \mu^3},
\label{zero-T}
\ee
in agreement with ref.~\cite{kap-toi}. However, 
if $rT \geq 1$, then, as we mentioned above, the screened potential
decays exponentially similar to normal Debye screening with an important difference
that the screening mass does not contain the electromagnetic coupling, $e$. On the
other hand, the magnitude of the screened potential is proportional to $e^2$. 
So formally for $e=0$ the oscillating 
potential vanishes, while the Debye one tends to the vacuum Coulomb expression.

The ratio of the main term in the potential at $T\neq 0$ to that at $T=0$ 
is equal to
\be
\frac{U(r, T)}{U(r, T=0)} = \frac{4\pi r T e^{-2\pi r T}}{1-e^{-4\pi r T}}\,.
\label{U-of-T-to-T0}
\ee
It is always smaller than unity. i.e. the screening is weakest at $T=0$.

%dsp

Let us turn now to non relativistic limit, when $m_F \gg T$, $\mu-m_F \ll m_F$, 
and for simplicity
$\tilde\mu = \mu - m_F \gg T$. The calculations go along the same lines with 
evident modifications. The poles of the distribution function $f$
are located at
\be
q_n = \left[ (\mu^2 - m_F^2) + 2i \pi \mu T (1+2n)\right]^{1/2}
\approx \sqrt{2 m_F\tilde\mu}\, 
\left[ 1 + \frac{i\pi \mu T (1+2n)} {\mu^2 -m_F^2}\right]\,,
\label{q-n-nonrel}
\ee  
%where $\tilde \mu = \mu - m_F$.

\begin{figure}[htbp]
	\begin{center}
	 \includegraphics[scale=0.65]{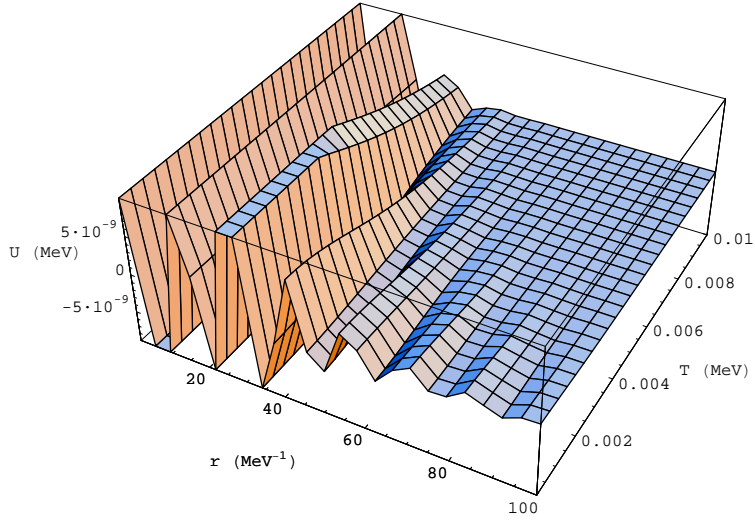}	
	\caption{{\it Friedel oscillations for massive fermions - 
see eq. (\ref{U-cut-fin}) -  
	with $m_F = 0.5 $ MeV, $\mu_F = 0.55$ MeV. Temperatures are in MeV
	and distances in MeV$^{-1}$. 
	The exponential damping
	at large distance and/or temperature, as well as oscillations 
	as a function of the distance $r$, can be seen.}}
	\label{fig_fer}
	\end{center}
	\end{figure}

The logarithmic singular part of $\Pi_{00}$ corresponding to this pole is
given by the same eq.~(\ref{Pi-n}) and the discontinuity on the cut is 
given by eq. (\ref{Delta-Pi}). An essential difference now is that the discontinuity
does not vanish near the branch point, $(4E_n^2 - 4q_n^2)=4m_F^2 \neq 0$:
\be 
\Delta \Pi_{00} \approx e^2 T m_F^2 /k.
\label{Delta-Pi-nonrel}
\ee
Thus the contribution of the $n$-th pole into the screened potential is equal to:
\be
U_n(r) = \frac{e^2 Q T m_F^2}{\pi^2 r}\, {\cal I}m\,\int_0^\infty
\frac{i dy \exp (2iq_n r - yr )}{\left[ k^2 + \Pi_{00}^{(n)+}(k)\right]
\left[ k^2 + \Pi_{00}^{(n)-}(k)\right]}\,.
\label{U-cut-n-B}
\ee
Here, as in the relativistic case above, $k = 2q_n + iy$. Neglecting $k^2$ in comparison
with $\Pi_{00}$, see eq. (\ref{Pi-0-rel}) and discussion below eq. (\ref{q-n-rel}),
we obtain:
\be
U_n(r) = \frac{Q e^2 T m_F^2}{ 16 \pi^2 q_n^4 r^2}\, {\cal I}m\,
\left[ i e^{2iq_n r} \right]=
\frac{Q e^2 T}{ 64 \pi^2 r^2 \tilde \mu^2}\,\cos (2\sqrt{2 m_F \tilde\mu} r) 
\exp\left[-2\pi (2n+1) \frac{ r T \mu}{\sqrt{2 m_F \tilde \mu}}\right].
\label{U-n-rel-1}
\ee
If temperature is not extremely small, the term with $n=0$ gives the slowest 
decreasing part of the potential, but for $T\rar 0$ we need to take into account
the whole sum $U_{cut}(r) = \sum U_n(r)$:
\be
U_{cut}(r) = \frac{Q e^2  T m_F^2} { 64\pi^2 r^2  \tilde\mu^2}\,
\cos \left(2\sqrt{2m_F \tilde\mu} \,r\right) \,
\frac{\exp \left( - \pi r T \sqrt{2m_F / \tilde \mu} \right)}
{1 - \exp \left( - 2\pi r T \sqrt{2m_F /\tilde \mu} \right)}\,.
\label{U-cut-fin}
\ee
Asymptotically for large $r$ but $2 \pi r T \sqrt{2m/\tilde \mu} < 1$ the 
potential tends to
\be
U_{cut}(r) = \frac{Q e^2 m_F \cos (2q_F r)}{64 \pi^3 r^3 q_F^3}\,, 
\label{U-cut-asym}
\ee
where $q_F = \sqrt{2\tilde\mu m_F}$.
The result agrees with that presented in ref.~\cite{kap-toi}.
The potential in eq. (\ref{U-cut-fin}) is plotted in fig. \ref{fig_fer} as a 
function of distance $r$ and temperature $T$ for $m_F = 0.5$ MeV and 
$\mu_F = 0.55$ MeV. 
Temperatures vary from $10^{-4}$ MeV and $10^{-2}$ MeV, which corresponds to 
$(1.16\cdot 10^{6} - 1.16\cdot 10^{8}) $~K. 
Distances vary from $1$ MeV$^{-1}$ to $100$ MeV$^{-1}$, 
corresponding to $(2\cdot 10^{-11} - 2\cdot 10^{-9})$ cm. 
The main features for the plot in the relativistic case is similar to the 
non-relativistic one.

Note in conclusion that above we have
neglected $\Pi_{00}$ in comparison with $4q_n^2$. It is 
justified for sufficiently small $e^2$. Otherwise one has to calculate the integral more
accurately taking into account the mild logarithmic singularity in $\Pi_{00}$ which
goes to infinity at the branch point for non-relativistic fermions
and goes to zero for relativistic ones.

%----------------------------------------------------------------------------------------------------------------------------------------
\section{Screening in bosonic plasma \label{s-bosonic}}
\label{bos-sec}
%----------------------------------------------------------------------------------------------------------------------------------------

As we have already mentioned the photon polarization tensor in presence 
of Bose condensate is infrared singular, having at small $k$ 
form~(\ref{low-k-Pi-00}).
The terms $\sim 1/k^2$ have been found in refs.~\cite{dlp,gr-3},
while $1/k$-term, which vanishes at $T=0$, has been found in ref.~\cite{dlp}.
Because of $1/k^2$ term the pole of the photon Green's function shifts from imaginary
axis in contrast to the usual Debye case when the pole is 
purely imaginary. Due to its real part the screened potential acquires an
oscillating factor superimposed on the exponential decrease~\cite{gr-2,dlp}.
The positions of poles in integral (\ref{U-of-r}) are given by the
equation $k^2 + \Pi_{00} (k) = 0$, which is convenient to write as:
\be
k^2 + e^2\left(m_0^2 + \frac{m_1^3}{k} + \frac{m_2^4}{k^2}\right) = 0,
\label{eq-pole-gnr}
\ee
where
\be
m_0^2 &=& \frac{C}{(2\pi^3) m_B} +
% \frac{T^2}{3} + 
%\frac{\zeta(3/2)(m_B T^3)^{1/2}}{(2\pi)^{3/2}} + 
h(T) + m_D^{(F)2} (T ,\mu_F) \label{m0-2},\\
m_1^3 &=& \frac{m_B^2T}{2} \label{m1-3},\\
m_2^4 &=& \frac{4m_B C}{(2\pi)^3}, \label{m2-4}
\ee
where $h(T)$ is defined in eq. (\ref{h-of-T}) and
$m_D^{(F)}$ is the fermionic Debye mass. For relativistic fermions it is given
by eq.~(\ref{Pi-0-rel}) and for non-relativistic ones by eq.~(\ref{m-D-nonrel}).
If plasma is electrically neutral because of the mutual compensation of bosons and fermions,
the chemical potential of fermions is expressed through
the amplitude of Bose condensate and $\mu_B = m_B$. % as follows:
%\be
%\mu_F \equiv \mu = [please \,\,\,find]
%\label{mu-F}
%\ee
However, one can imagine the case when there are two types of charged bosons
and neutrality is achieved by the opposite charge densities of these bosons. In such
plasma the fermionic Debye mass is zero.

In what follows we analyze different contributions to the electrostatic potential $U(r)$
for different limiting values of the parameters.
In sec. \ref{subsec_poles} we investigate further the contribution from the 
poles in integral (\ref{U-of-r}).
In sec. \ref{subsec_im_axis} we present the contribution from the imaginary axis 
which arises when 
the integrand in eq. (\ref{U-of-r}) is not an even function of $k$. 
Finally in sec. \ref{subsec_branch_cuts} we calculate the contributions 
from integration along the
branch cuts of the logarithmic terms in $\Pi_{00}$, see eq.~(\ref{Pi-00-B}).
The integration contour is similar to that for fermions,
fig.\ref{fig_contour} but the positions of the poles are evidently shifted,
see the following subsection. 

%-----------------------------------------------------------------------------------------------
\subsection{Contribution from poles}
\label{subsec_poles}
%-----------------------------------------------------------------------------------------------
At low temperatures the four roots of 
eq. (\ref{eq-pole-gnr}) are given by:
\be
k_{1,2,3,4} = \pm \frac{i}{\sqrt{2}} \left[e^2m_0^2 \pm \sqrt{e^4 m_0^4 - 4e^2 m_2^4} 
\right]^{1/2}.
\label{k-1-4}
\ee
As is mentioned above, we are interested only in the poles in the first quadrant 
in the complex $k$-plane.
If $e^4 m_0^4 > 4e^2 m_2^4$, all the poles are purely imaginary and the Coulomb
potential is screened exponentially, similar to the usual Debye situation. The poles
on the positive imaginary axis are situated at
\be
k_{1,2} = \frac{i e m_0}{\sqrt{2}} \left( 1 \pm \sqrt{ 1 - 4m_2^4/e^2 m_0^4 }\right)^{1/2}.
\label{k-1-2}
\ee
The contribution of these poles into the potential is
\be
U(r) = \frac{Q }{4\pi r}\, \frac{ k_1^2 e^{ik_1r} - k^2_2 e^{ik_2r }}
{k_1^2 - k_2^2}.
\label{U-pole-12}
\ee
In the limit of small ratio $m_2^2 /e m_0^2$ the potential becomes:
\be
U(r)_{pole} \approx \frac{Q}{4\pi r}\,\left[ \exp\left(-e m_0 r \left(1-\frac{m_2^4}{2e^2m_0^4}\right) \right) -
\frac{m_2^4}{e^2 m_0^4}\,\exp\left(-m_2^2 r/m_0\right)\right].
\label{U-m2}
\ee 
Thus for a small $m_2$ the screening, though exponential, can be much weaker than
the usual Debye one.

In the opposite case, $e^4 m_0^4 < 4e^2 m_2^4$, the poles acquire real part and now only one pole is situated in the first quadrant. 
The potential oscillates around the exponentially decreasing 
envelope~~\cite{gr-2,dlp}. 
The result is especially simple in the limit of large $m_2$:
\begin{eqnarray}
U(r)_{pole} = \frac{Q}{4\pi r}\, \exp \left( - \sqrt{e/2} m_2 r\right) 
\cos \left(\sqrt{e/2} m_2 r\right).
\label{U-j}
\end{eqnarray}
More interesting situation is realized at larger temperatures, when the
term $m_1^3/k$  in the polarization operator, eq. (\ref{eq-pole-gnr})
is non-negligible. 
The contribution of the poles into the asymptotics of the screened potential
is similar to the above considered case of low $T$ if $m_2$ dominates in
$\Pi_{00}$, but for a small $m_2$, e.g. if $C=0$, the poles are situated
at $k= e^{2/3} (-1)^{1/3}(m_B^2 T/2)^{1/3}$. The potential exponentially decreases
at large distances but the power of the exponent is proportional to temperature
and at small $T$ the decrease of $U(r)$ may be rather weak.

%-----------------------------------------------------------------------------------------------
\subsection{Contribution from the integral along the imaginary axis}
\label{subsec_im_axis}
%-----------------------------------------------------------------------------------------------

Because of the odd term, $m_1^3/k$, in the polarization operator 
the imaginary part of 
integral~(\ref{U-of-r}) along the imaginary axis in the complex $k$-plane is non-zero
and the screened potential drops as a power of $r$:
\begin{eqnarray}
U(r) = -\frac{Q e^2 m_1^3}{2\pi^2 r^2}\,\int_0^\infty \frac{dz \exp(-z)} 
{\left[- (z/r)^2 + e^2  (m_0^2 - m_2^4 r^2/z^2 )\right]^2 + e^4 m_1^6 r^2/z^2}.
\label{power-U}
\end{eqnarray}
The previous expression has been obtained by substituting $k=iy$ and then $z=yr$.
If $m_2 \neq 0$ the dominant term at large $r$ behaves as
\be
U(r) = -\frac{ 12 Q m_1^3}{\pi^2 e^2 r^6 m_2^8}. 
\label{U-of-r-C0}
\ee

However, if  the temperature is not zero and the bosonic chemical potential 
reaches its upper limit, $\mu = m_B$, but the condensate is not yet formed,  
the term proportional to $m_1$ dominates and the asymptotic decrease of the
potential becomes much slower:
\be
U(r) = -\frac{ Q }{\pi^2 e^2 r^4 m_1^3} =  -\frac{2 Q }{\pi^2 e^2 r^4 m_B^2 T}\,.  
\label{U-of-r-m1}
\ee
So the formation of the condensate manifests itself by a 
strong decrease of screening. This effect 
may be a signal of formation of Bose condensate.

It is interesting that the screened potential is inversely proportional
to the fine structure constant $\alpha = e^2/4\pi$.

%-----------------------------------------------------------------------------------------------
\subsection{Contribution from the logarithmic branch cuts}
\label{subsec_branch_cuts}
%-----------------------------------------------------------------------------------------------

Let us estimate now the effects of the logarithmic singularities of $\Pi_{00}$
on the asymptotics of the screened potential (analogue of the Friedel oscillations). 
Technically the calculations are
similar to those made in sec.~\ref{s-friedel} but the results are noticeably 
different. We assume here that the chemical potential of bosons reaches 
its maximum value,
$\mu = m_B$. For smaller $\mu$ there is not much difference between bosons and
non-degenerate fermions, while for $\mu = m_B$ new phenomena arise, which
are absent for fermions.

The poles in the integrand of eq. (\ref{Pi-00-B}), which lead to the 
singularities of $\Pi_{00} (k)$ in the first quadrant of the complex $k$-plane,
are situated at
\be
q_n = \left( 4 i \pi n T m_B\right)^{1/2} \left( 1 + i\pi n T/m_B \right)^{1/2}. 
\label{qn-B}
\ee
Here $n$ runs from 1 to infinity, because there is no pole at $q=0$
since the numerator of the integrand is proportional to $q^2$.

The singularities in $\Pi_{00} (k)$ are situated at such $k$ where 
the singularities
of the integrand in eq. (\ref{Pi-00-B}) pinch the integration contour, i.e. 
as above, at $k_n =  2q_n$. The singular part of $\Pi_{00}$ is calculated
in the same way as it has been done for fermions and is equal to the residue 
of the integrand:
\be
\Pi_{00}^{(n)B} = -\frac{ie^2 T E_n^2}{2\pi k}\,\ln \left(
\frac{k-2q_n - i\epsilon}{k+2q_n + i\epsilon}
\right)\,,
\label{Pi-n-B}
\ee  
where $E_n = \sqrt{q_n^2 + m_B^2}$. 

The discontinuity of this term across the logarithmic cut is 
$\Delta \Pi_{00}^{(n)B} = e^2 T E_n^2 /k$. Correspondingly the contribution of this
singularity into the asymptotics of $U(r)$ is given by:
\be
U_n^{B} (r) = -\frac{Qe^2 T}{2\pi^2 r}\,{\cal R}e\, \int_0^\infty
\frac{ dy E_n^2 e^{2i q_n r} e^{-ry} }{\left[k^2 + \Pi_{00}^{(+)}\right]
\left[k^2 + \Pi_{00}^{(-)}\right]}\,,
\label{U-n-B}
\ee
where $k = 2q_n + iy$ and $E_n^2 = q_n^2 + m_B^2$, and $\Pi_{00}^{\pm}$ are the
values of the polarization tensor on right and left banks of the cut. Note that
at $r \rar \infty$ the effective $y$ is small, $y\sim 1/r$.

An important difference between bosonic and fermionic cases is that the
position of the pole for fermions, 
eqs. (\ref{q-n-rel},\ref{q-n-nonrel}), does not
move to zero when $T\rar 0$, while for bosons $q_n^2 \sim T$. Correspondingly
one can neglect $\Pi_{00}^F$ in comparison with $k_n^2$, while it may be an 
invalid approximation for bosons.

\begin{figure}[htbp]
	\begin{center}
	 \includegraphics[scale=0.55]{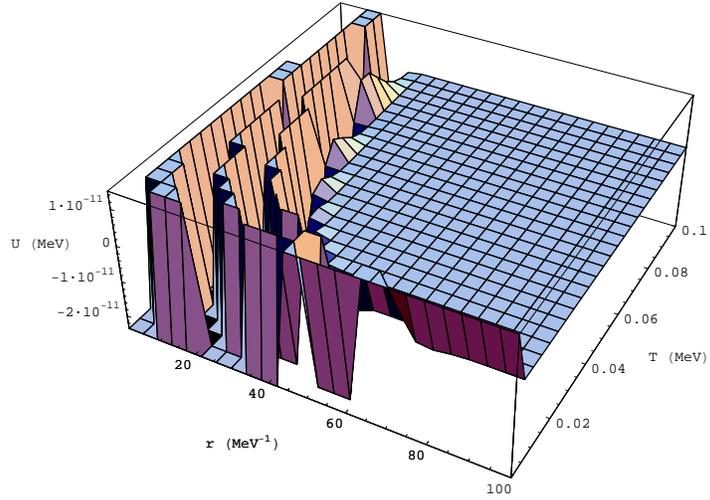}	
	\caption{{\it Oscillation of the electrostatic potential in 
presence of bosonic plasma,
	see eq. (\ref{U-1}). 
	The boson mass is equal to the fermion one 
in Figure \ref{fig_fer}, $m_B = 0.5 MeV$
	and the chemical potential is $\mu_B = m_B$. 
	Temperatures are in $MeV$
	and distances in $MeV^{-1}$.
	}}
	\label{fig_bos}
	\end{center}
	\end{figure}

Let us first consider the case of low temperatures when $\Pi_{00}$ is 
dominated by the constant fermionic contribution, $\Pi_{00}^F \approx m_D^2$,
where $m_D^2$ is given either by eq. (\ref{Pi-0-rel}) or (\ref{Pi-0-nonrel}).
At large $r$ and non-zero $T$ the logarithmic contribution into the
screened potential is essentially given by the first term with $n=1$:
\be
U_1 (r) = -\frac{Q\pi^2}{2e^2}\,\frac{T m_B^2}{r^2 \mu_F^4}\,
\exp \left(-2\sqrt{2\pi m_B T} r\right)\,\cos\left(2\sqrt{2\pi m_B T} r\right).
\label{U-1}
\ee
Here we took the relativistic limit for $\Pi_{00}^F$. The result is easy to 
rewrite in non-relativistic case.
The potential in eq. (\ref{U-1}) is plotted in figure \ref{fig_bos}. 
The bosonic chemical potential is taken to be equal to its limiting value,
$\mu_B = m_B$, and
the boson mass is assumed to be the same as 
the fermion mass in fig. \ref{fig_fer}, $m_B = m_F = 0.5$ MeV. Such a low 
mass of bosons is chosen simply for illustration. In realistic case charged
bosons are much heavier than the charged fermions, though it is not
excluded that there exists an unknown gauge symmetry with charged bosons
lighter than fermions. 

The temperature in fig. \ref{fig_bos}
varies from $10^{-4}$ MeV to  $0.1$ MeV, corresponding to 
$(1.16\cdot 10^{8} - 1.16\cdot 10^{9})$ K, while distances vary from 
$1$ MeV$^{-1}$ to $100$ MeV$^{-1}$, corresponding to 
$(2\cdot 10^{-11} - 2 \cdot 10^{-9})$ cm.

\begin{figure}[htbp]
	\begin{center}
	 \includegraphics[scale=0.6]{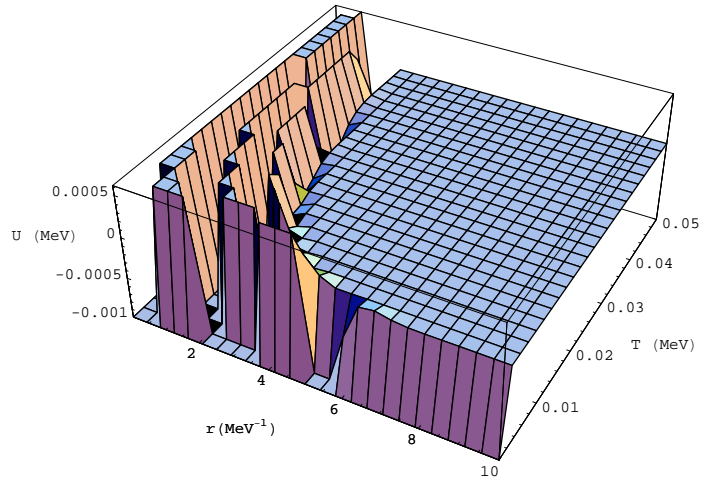}	
	\caption{{\it Oscillation of the electrostatic potential in 
presence of bosonic plasma
	- see eq. (\ref{U-1}). 
	The boson chemical potential is equal to its mass, 
$\mu_B = m_B = 100 MeV$. 
	Temperatures are in $MeV$
	and distances in $MeV^{-1}$.
	In the picture are evident the oscillations due to both the 
temperature $T$ and the
	distance $r$ as well as the exponential damping in both the directions.
	}}
	\label{fig_bos_100}
	\end{center}
	\end{figure}

Figure \ref{fig_bos_100} shows the same potential but with higher mass for 
bosons, $m_B = 100$ MeV, that is of the order of the pion mass. The fermion 
mass and chemical potential are taken the same as above.
The temperature varies in the range $10^{-6} - 5\cdot 10^{-2}$ MeV
or $1.16\cdot 10^4 - 5.8\cdot 10^8$ K
and the distance in $10^{-2} <r({\rm MeV})^{-1} <10$ corresponding to
$2 \cdot 10^{-13} < r ({\rm cm}) < 2 \cdot 10^{-10}$.
We can see from these figures that if we increase the boson mass, 
the bosonic potential fades away faster.


In the limit of $T\rar 0$ (analogous to the discussed above Friedel case)
we should take the sum $\sum_{n=1}^{\infty} U_n$ because all the terms 
are of the
same order of magnitude and $n_{eff} \sim 1/(4\pi m_B T r^2)$. So we 
could expect that
the sum is inversely proportional to $T$ and the potential is non-vanishing
at $T=0$, the same as in the fermionic case. However, the summation is not so 
simple as previously because we do not deal now with geometric progression,
$\exp (-a n)$ but with more complicated function, $\exp (-b\sqrt{n})$. Since
the effective values of $n$ are big, we can express the sum as an integral
and obtain, in the leading approximation $\Pi_{00} =  m_D^2$,
that the potential is proportional to the temperature $T$ and hence vanishes:
\be
U(r)^B = 
- \frac{Q T \pi^2}{2 e^2 r^2 \mu_F^4}
{\cal R}e \sum_{n=1}^\infty E^2_n e^{2iq_nr} \approx 
- \frac{Q T \pi^2}{2 e^2 r^2 \mu_F^4}
{\cal R}e\, \int_1^\infty dn \, E_n^2 e^{2iq_nr} \sim T .
\label{U-B-fr1}
\ee
The real part of the integral $ \int_1^\infty dn \, e^{2iq_nr} $ 
written above is equal to:
	\begin{eqnarray*} 
	\frac{\exp (-2\sqrt{2\pi Tm}r)}{4r} 
	\left[\sqrt{\frac{2}{\pi mT}} 
	\left( \cos (2\sqrt{2\pi Tm}r)- \sin (2\sqrt{2\pi Tm}r)
	\right)
	- \frac{1}{2 \pi mrT} \sin (2\sqrt{2\pi Tm}r)
	\right],
	\end{eqnarray*}
that goes to the constant value $-1$ in the $T \rar 0$ limit. So the 
whole expression in eq. (\ref{U-B-fr1}) is proportional to $T$.

It is important to stress that the previous result is valid in the 
limit  $T m_B r^2 \ll 1$, which
 means that it is applicable at small distances $r_B \ll 1/\sqrt{m_B T}$.
On the other hand at large distances $r$ and non vanishing $T$ 
one should consider the 
expression in eq. (\ref{U-1}) which is similar to the fermionic 
Friedel term in eq. (\ref{U-cut-fin})
but has different dependence on the coupling constant $e$ since 
it goes like $e^{-2}$, while Friedel oscillations go like $e^2$. 
Hence we have non analytic dependence 
on the coupling constant $e$ in presence of bosons. Similar dependence 
on $e^{-2}$ was found in sec. \ref{subsec_im_axis}.

There are also differences arising from the fact that in 
the limit $T \rar 0$ the
poles of the boson distribution function go to zero, see eq. (\ref{qn-B}), 
while the poles of 
the fermion distribution function tend to the non vanishing value 
$q_F$, see eq. (\ref{q-n}).
Hence Friedel oscillations for fermions start from their 
maximum amplitude at $T=0$ and then
exponentially decrease with temperature, while for bosons the 
effect vanishes at $T=0$, 
then linearly increases with $T$ and finally exponentially decreases.
Another consequence is that the argument of the oscillating cosine function 
depend on $T$ for bosons but not for fermions. Hence the boson potential 
does not oscillate
at small temperatures. 

At high fermionic chemical potential $\mu_F$ and small temperature $T$, 
the boson oscillations typically go to $0$ at smaller distances than the 
fermionic ones, which
are observable at distances $r \leq T$. 
On the other hand lowering the boson mass $m_B$ the exponential 
damping is weaker
but at the same time oscillations fade away.

If the condensate is formed, $\Pi_{00} $ would be dominated 
by the singular term
$e^2 m_2^4/k^2$ and according to eq. (\ref{U-n-B}) the contribution of $n$-th
branch point into the screened potential becomes:
\be
U_n^B(r) = - \frac{Q T m_B^2}{2\pi^2 e^2 m_2^8 r^2}\,
{\cal R}e\left[k_n^4 e^{ik_n r} \right].
\label{U-n-C}
\ee
Again, at large $r$ and non zero $T$ the $n=1$ term is dominant. It 
oscillates and exponentially decreases according to eq. (\ref{U-1}).
However, the sum ${\cal R}e\sum_n U_n^B$ vanishes as above, eq. (\ref{U-B-fr1}). 
Probably the vanishing of $U^B(r) $ at small $T$
in the leading order is a more general feature. At least
the sub-leading (at small $T$) terms in $k_n $ and in $\Pi_{00}$ vanish as well.
If we take into account the imaginary part of $\Pi_{00}$ due to the logarithmic
cut, the result still remains proportional to a power of
temperature after summation. On the other hand, as
we see below, in absence of condensate
the potential not only survives at $T \rar 0$ but rises as an inverse power of $T$.

Let us turn now to a more interesting though probably less realistic case when
fermions are absent in the plasma, chemical potential of bosons is maximally allowed,
$\mu_B = m_B$ but the condensate is not formed. 
In the standard model a neutral system has necessarily a fermionic component because 
fermions are lighter then bosons. Anyway we can imagine systems where the 
electric charge is compensated by other heavier bosons which do not 
condense or models with extra $U(1)$ sector and different particle content. 
In this situation fermions may be absent.
Under these conditions $\Pi_{00}$ vanishes
when $T\rar 0$. The position of the branch points of the 
logarithm $k_n =2q_n$ also
tend to zero and the screening due to logarithmic discontinuity 
may be non-vanishing at
$T=0$. Indeed, let us turn again to eq. (\ref{U-n-B}). The integral 
goes along the
contour $k= k_n + iy$ and $y\sim 1/r$ is very small. 
We assume that $r>1/\sqrt{T m_B}$. 
Thus $k^2 \approx k_n^2 = 16 i \pi n T m_B$. 
Let us now estimate $\Pi_{00}$ at $k=k_n$.
At small temperatures, 
when $z^2 \equiv (E_B - m_B)/T \approx q^2/(2m_BT)$, $\Pi_{00}$ can be
presented as:
\be
\Pi_{00}(k) = \frac{e^2 m_B^2 T}{\pi^2 k} \int \frac{dz z}{\exp(z^2) -1}\,
\ln \bigg| \frac{\sqrt{8 m_B T} z +k}{\sqrt{8 m_B T} z -k} \bigg|.
\label{Pi-00-of-k}
\ee 
Notice in passing that if $k< \sqrt{8m_BT}$, then $\Pi_{00}$ 
behaves as $m_1^3/k$
in agreement with eqs. (\ref{eq-pole-gnr},\ref{m1-3}), while at large $k$,
$k> \sqrt{8m_B T}$, it has the following asymptotic behavior:
\be
\Pi_{00}(k) \approx \frac{\sqrt{2} 
e^2 m_B^{5/2} T^{3/2}\zeta(3/2)}{\pi^{3/2} k^2},
\label{Pi-large-k}
\ee 
where $\zeta(3/2) \approx 2.6$.
The singular part of $\Pi_{00}$,
eq. (\ref{Pi-n-B}), at $k= k_n + iy$ is equal to
\be
\Pi_{00}^{(+)}(k_n+iy) = - \frac{i^{1/2} e^2 T^{1/2} m_B^{3/2} }
{8 \pi^{3/2} n^{1/2}}\,
\left[\ln (y/8 \sqrt{\pi n m_BT}) + i\pi/2 \right].
\label{Pi-of-kn-sing}
\ee
For $\Pi_{00}^{(-)}$ the last factor is changed to 
$\left(\ln y /8 \sqrt{\pi n m_BT} - 3i\pi/2 \right)$.
The factor in the denominator of the logarithm comes from 
$|k + 2 q_n| = 4|q_n|$ in eq. (\ref{Pi-n-B}).

The screened potential (\ref{U-n-B})
at large distances, i.e. for $ 8\pi T m_B r^2 > 1$, is dominated by $n=1$. 
One can check that
$|\Pi_{00}(k_1)| > |k_1^2|$, so the latter can be neglected in the 
denominator of
eq. (\ref{U-n-B}). Keeping in mind that we will use the result below for 
arbitrary $n$
for which $|\Pi_{00}(k_1)| > |k_1^2|$, we write:
\be
U_n(r) \approx \frac{32\pi Q n }{e^2 m_B r^2}\, {\cal R}e \left[ i e^{2i q_n r} \,
\int_0^\infty \frac{dx e^{-x}}{\ln^2 (x / 8 \sqrt{m_B\pi n T}r) - 
i \pi\ln (x / 8 \sqrt{m_B\pi nT}r) + 3\pi^2/4}\right],
\label{U1-log}
\ee
where $x = yr$. For large logarithm the leading 
part of the integral can be approximately evaluated 
leading to the result:
\be
U_1 (r) = - \frac{32\pi Q}{e^2 m_B r^2} \frac{e^{-2\sqrt{2\pi Tm_B} r}}
{\ln^2(8 \sqrt{\pi m_B T } r) }\,
\sin (2\sqrt{2\pi T m_B }r)\,.
\label{U1-large-r}
\ee
Note that $U_1(r)$ is inversely proportional  to the electric charge and 
formally vanishes at $T \rar 0$, but remains finite if $\sqrt{T m_B} r$ 
is not zero.

For smaller distances, or such small temperatures that $8\pi Tm_B r^2 \ll 1$,
all $n$ up to $n_{max} \sim 1/(8\pi T m_B r^2) $ make comparable 
contributions. Thus
we have to sum over $n$. If $n_{max} \gg 1$ the sum can be 
evaluated as an integral
over $n$. Now, for large $n$, $k_n^2\sim n$ and may be 
comparable to 
$\Pi_{00}(k_n)$ which, according to eq. (\ref{Pi-of-kn-sing}), 
drops as
$1/\sqrt{n}$.
$\Pi_{00}(k_n)$ would be smaller by magnitude than $k_n^2$ for
\be
n> n_0 \approx 10^{-3} \left(m_B/T\right)^{1/3} 
\ln^{2/3} \left(\sqrt{ 8m_B Tr^2}\right).
\label{n-0}
\ee
This condition makes sense if $n_0 < n_{max}$ or 
$r \ln^{1/3} (\sqrt{8m_BTr^2}) < 5/(T m_B^2)^{1/3} $.
For larger $r$ we return to domination of $\Pi_{00}$.
We should check that the condition 
\be
r\ln^{1/3} (\sqrt{8m_BTr^2}) > 5/(T m_B^2)^{1/3}
\label{r-min}
\ee
does not contradict the condition of large $n_{max} $. The latter reads 
\be
r < 1/\sqrt{8\pi T m_B}.
\label{r-max}
\ee
If we neglect the logarithmic factor, both conditions would be compatible for 
$T/m_B < 4 \cdot 10^{-9}$. 
Thus both cases of dominant $\Pi_{00} (k_n)$ or $k_n^2 $ can be 
realized depending 
upon relation between $r$, $T$, and $m_B$.

Let us consider smaller temperatures when $|\Pi_{00} (k_n)| > |k_n^2|$. 
The potential
in the limit of small $\pi T m_B r^2 $ is equal to
\be
U(r) = \frac{32 \pi Q}{e^2 m_B r^2}
{\cal I}m \left[ \sum_n n e^{2iq_n r} 
\frac{\ln^2 (\sqrt{8m_B T } r) +i\pi \ln (\sqrt{8m_B T } r) +
3\pi^2/4 }{\left( \ln^2 (\sqrt{8m_B T } r) + 3\pi^2/4 \right)^2 + 
\pi^2 \ln^2 (\sqrt{8m_B T } r) } \right].
\label{bos-fr-U}
\ee
Since the sum
\be
\sum_n  n e^{2iq_n r}  \approx  2\int d\eta \, \eta^3 e^{4 i\sqrt{i\pi T m_B} r \eta} \approx
-\frac{12}{256 \pi^2 T^2 m_B^2 r^4},
\label{sum-n}
\ee
where $\eta = \sqrt{n}$, is real in leading order in $1/(16\pi T m_B r^2)$, a non-vanishing contribution comes from the 
imaginary part of the numerator of the integrand and we obtain for the analogue of Friedel
oscillations in purely bosonic case:
\be
U(r) \approx -\frac{3 Q}{2 e^2 T^2 m_B^3 r^6 \ln^3 \left(\sqrt{8m_B T } r\right)}.
\label{U-B-fin}
\ee
The result has some unusual features. First, the potential decreases monotonically without
any oscillations. Second, it is inversely proportional to the temperature, so 
the smaller is $T$, the larger is the potential. However, the effect exists 
for sufficiently small $r$,
$r \ll 1/\sqrt{16 \pi T m_B}$, i.e. if $T= 0.1 $K and $m_B = 1 GeV$ the 
distance should be bounded from above as 
$r \ll 3\cdot 10^{-8}$ cm. Another obstacle to realization of such screening
behavior is that with fixed charge asymmetry Bose particle should condense and the dominant
term in $\Pi_{00}$ becomes $4 m_B C /(2\pi)^3$. In this conditions we arrive to potential
(\ref{U-n-C}) which vanishes at $T= 0$.

%----------------------------------------------------------------------------------------------------------------------------------------
\section{Conclusion \label{s-conclusion}}
\label{conclusion}
%----------------------------------------------------------------------------------------------------------------------------------------

We have calculated the electrostatic potential between two test charges in plasma with
electrically charged bosons and fermions. The new part of our consideration is an inclusion of
the effects of the Bose condensate into the screening phenomena in plasma. To this end the chemical
potential of bosons is taken equal to the maximally allowed value, that is to the boson
mass,  $\mu= m_{B}$. In this case the bosonic contribution to the time-time component of the photon
polarization operator in plasma, $\Pi_{00} (k)$,
acquires an infrared singular contribution proportional to $T/k$ even
before formation of the condensate and $1/k^{2 }$ after formation of the condensate. Such terms
drastically change the form of the screened potential $U(r)$.

All the calculations have been done in the lowest order in the electromagnetic coupling constant,
$e$. We have imposed the condition of electric neutrality of the plasma, assuming that
bosons and fermions compensate each other charge. We have noticed however, that the
screened potential demonstrates a very interesting and unusual behavior  as a function of temperature
if fermions are absent. Such a situation cannot be realized in realistic equilibrium plasma because
the lightest charged fermions (electron/positrons) are lighter than charged bosons. However, one can
imagine a hypothetical case of a new gauge $U(1)$-symmetry with lighter charged bosons.

We started with purely fermionic plasma and reconstructed the known Friedel oscillations of $U(r)$
 both in relativistic and non-relativistic limits using somewhat different technique. 
 We obtained an explicit expression for $U(r)$ at non-zero
 temperature which, to the best of our knowledge, is absent in the literature. 

The main part of our work is dedicated to the new phenomena created by the singularities of 
$\Pi_{00}(k)$ in the complex $k$-plane. We reproduced the previously obtained results that 
due to $1/k^{2}$ term  the pole of the static photon propagator acquires non-zero real part and
because of that the screened potential oscillates with an exponentially decreasing envelope. 

At non-zero temperature and $\mu=m_{B}$ the polarization operator obtains an odd
contribution with respect to the transformation $k\rar -k$ and because of that the integral 
along the imaginary axis in the
complex $k$-plane, which determines the asymptotic behavior of $U(r)$, becomes non-vanishing.
It leads to monotonic power law screening $U\sim 1/r^{6}$, eq. (\ref{U-of-r-C0}) if the condensate 
has not yet been formed. After the condensate formation the screened potential behaves as 
$U(r) \sim 1/(e^{2}r^{4}T)$, eq. (\ref{U-of-r-m1}). 
Such a change in the screening may be a signal of  the condensate formation.

We have also considered an analogue of the Friedel oscillations in the bosonic case.
The origin  of the phenomenon is the same as in the fermionic case but the resulting potential is
quite different. The Friedel oscillations can be understood as a result of pinching the integration
contour in the complex $k$-plane by the logarithmic branch point of $\Pi_{00}$ and the poles of
the bosonic (or fermionic) distribution functions. However, the poles of bosonic distribution moves 
to zero when temperature tends to zero, while the fermionic ones keep a finite value. This leads to 
completely different behavior of the potential as a function of temperature. The potential vanishes
when $T$ goes to zero for mixed bosonic and fermionic plasma. In the case that it is dominated 
by the  first pole, for large $r$ and non-zero $T$, it goes as in 
eq. (\ref{U-1}) and at small $T$ the exponential screening is quite mild.
For purely  bosonic plasma the ``Friedel'' part of the screening is  given by eqs.
(\ref{U1-log}) and (\ref{U-B-fin}).  If $Tm_{B} r^{2}$ is not small the potential oscillates and
exponentially decreases, while for smaller $T$ 
it does not oscillates and is proportional to 
$1/(e^{2}T^{2})$. The $1/e^{2}$ behavior looks puzzling  but one should remember that it is
an asymptotic result for large distances. However, if we take the formal limit $e\rar 0$ the screening 
would disappear together with $e$. Similar reasoning is applicable to $1/T^{2}$ behavior: this is
true only for large but simultaneously sufficiently small distances $r<1/\sqrt{16\pi m_{B}T}$, when
the $k^{2}$ part of the photon Green's function is sub-dominant.

We see that the screening is quite different in different limits and it would be very interesting
to study this rich behavior experimentally.

%----------------------------------------------------------------------------------------------------------------------------------------
{\vspace{0.8cm} \noindent \bf Acknowledgments}
%----------------------------------------------------------------------------------------------------------------------------------------
G. P. acknowledges support provided by the grant 
program PASPA-UNAM and PAPIIT-UNAM, number IN112308.

%----------------------------------------------------------------------------------------------------------------------------------------

\end{document}